\documentstyle[12pt]{article}
\begin{document}

\begin{titlepage}
\noindent BUHEP-93-15 \hfill July, 1993\\
\noindent Wash.\ U.\ HEP/93-34\\
\begin{center}

{\Large\bf A Lattice Calculation of the Isgur-Wise Function}

\vspace{1.5cm}
{\bf C. Bernard$^{(1)}$, Y. Shen$^{(2)}$ and A. Soni$^{(3)}$}\\
\end{center}
\vspace{1.0cm}
\begin{flushleft}
{}~~$^{(1)}$Physics Department, Washington University, St. Louis, MO 63130, USA
\\

{}~~$^{(2)}$Physics Department, Boston University, Boston, MA 02215, USA \\

{}~~$^{(3)}$Physics Department, Brookhaven National Laboratory, Upton, NY
11973, USA \\
\end{flushleft}
\vspace{1.5cm}

\abstract{
 We calculate the Isgur-Wise function ($\xi_0$)
by measuring the heavy-heavy meson
transition matrix element on the lattice
in the quenched approximation. The standard Wilson action is used
for both the heavy and the light quarks. Our numerical results
are compared with various model calculations and experimental data.
In particular, using a linear fit, we find $\rho^2 = 1.24\pm .26 \pm .26 $.
(The slope of $\xi_0$ at the zero-recoil point is $-\rho^2$.)
Using instead the parametrization of $\xi_0$ suggested by
Neubert and Rieckert and fitting the lattice data up to $v\cdot v' \sim 1.2$,
we find $\rho^2_{NR}=1.41\pm.19\pm.19$

}
\vfill
\end{titlepage}

\section{Introduction}

The physics of heavy quark systems has been one of the active research topics
in the last few years (for reviews, see \cite{Wise}).
Consider a meson system consisting of
a heavy quark $Q$ and a light quark ${\bar q}$. When $m_Q \to \infty$
the heavy quark essentially becomes a static color source. It can be described
by a simple heavy quark effective field theory (HQEFT) with flavor
and spin symmetries.
Because of these new flavor and spin symmetries in HQEFT, the meson decay
matrix element can be simplified and different decay processes can be related
to each other.
For example, in the case of $B \to D$ decay there are two
independent form factors in the full theory. However, if we assume both
B and D mesons are heavy enough, using HQEFT the number of unknown form factor
can be reduced to one. We have \cite{Wise}
\begin{equation}
<D(v^\prime) |{\bar c}\gamma_\nu b |B(v)> = \sqrt{m_B m_D} C_{cb}(\mu)
\xi_0 (v^\prime \cdot v; \mu) (v + v^\prime)_\nu ~,
\label{eq:BtoD}
\end{equation}
where $v_\nu, v^\prime_\nu$ are the four-velocity and $m_B, m_D$ are
the B and D meson mass, respectively. The constant $C_{cb}(\mu)$
comes from integrating the full QCD contribution from the heavy quark mass
scale
down to a renormalization scale $\mu \ll m_D$
\begin{equation}
C_{cb}(\mu) = \left [{\alpha_s(m_D)\over\alpha_s(m_B)}\right]^{6/(33-2N)}
\left [{\alpha_s(m_B)\over\alpha_s(\mu)}\right]^{a(v\cdot v^\prime)}~,
\end{equation}
where $a(v\cdot v^\prime) $ is a slowly varying function of
$ v^\prime \cdot v $ which vanishes at $v = v^\prime$ \cite{Wise}.
The Isgur-Wise function $\xi_0 (v^\prime \cdot v; \mu)$ represents the
interactions
between the light degrees of freedom in the heavy meson system.
It thus has to be calculated using nonperturbative methods.

On the lattice the heavy meson system can be studied in two different
approaches. One is to keep the heavy quark dynamical by using the standard
Wilson action. This may require extrapolation to the physical heavy meson
mass of interest.
The alternative is to integrate out the heavy quark first and
derive an effective action including only the light degrees of freedom and
then perform numerical simulation using
this effective action \cite{ogilvie}.
Here we stay
with the first approach (see \cite{Mack} for reviews
of the second approach).
We reported preliminary results from our approach
at the
{\it Lattice '92}
conference \cite{BSS}.

It is important to note that
using the flavor symmetry of HQEFT the Isgur-Wise function
relevant to the
$B \to D$ decay of Eq.\ (\ref{eq:BtoD}) can be obtained
also from the $D \to D$
elastic scattering matrix element \cite{Wise}
\begin{equation}
<D_{v^\prime} | {\bar c}\gamma_\nu c|D_{v}> = m_D C_{cc}(\mu)
\xi_0 (v\cdot v^\prime; \mu) (v + v^\prime)_\nu ~,
\label{eq:matrix}
\end{equation}
where
\begin{equation}
C_{cc}(\mu) = \left[{\alpha_s(m_c) \over \alpha_s(\mu)}\right]
^{ a(v\cdot v^\prime)}~.
\label{eq:cfactor}
\end{equation}
This of course requires that the D meson be sufficiently heavy for the onset
of the heavy quark limit (HQL).

Conventionally
$B \to D$ (here we use B and D as generic names for heavy
pseudoscalar mesons, they do not necessarily represent the physical B and
D mesons) transition matrix elements can be parametrized as
\begin{equation}
<D(p^\prime)|V_\nu|B(p)> = f_+(q^2) (p^\prime + p)_\nu + f_-(q^2)(p -
p^\prime)_\nu~,
\label{eq:form1}
\end{equation}
where $q^2 = (p^\prime-p)^2$ is the momentum transfer between the initial
and final states and $V_\nu$ is a vector current.
One can also use an equivalent form
\begin{equation}
<D(p^\prime)|V_\nu|B(p)> = f_+(q^2)\left[(p^\prime+p)_\nu - {m_B^2-m_D^2
\over q^2}
(p-p^\prime)_\nu\right] + f_0(q^2){m_B^2-m_D^2\over q^2}(p-p^\prime)_\nu~,
\label{eq:form2}
\end{equation}
with relation between the form factors $f_+, f_-$ and $f_0$
\begin{equation}
(m_B^2-m_D^2)f_0(q^2) = (m_B^2-m_D^2)f_+(q^2) + q^2 f_-(q^2)~.
\label{eq:fminus}
\end{equation}
For the elastic scattering process ($m_B = m_D$),
we have $f_-(q^2) = 0$,
which follows from current conservation.
Using the relation $v_\nu = p_\nu/m$,
Eq.\ (\ref{eq:form1}) becomes
\begin{equation}
<D(v^\prime)|V_\nu|D(v)> = m_D f_+(q^2) (v^\prime + v)_\nu ~.
\label{eq:forwd}
\end{equation}
Comparing this with Eq.\ (\ref{eq:matrix}) one finds the simple relation
between $f_+$ and $\xi_0$
\begin{equation}
f_+ = C_{cc} \xi_0 ~.
\label{eq:fplusxi}
\end{equation}
The lattice calculation method for $f_+$ has been well
established \cite{Bernard,ELC}.
We will use that method through equation (9)
and deduce $\xi_0$.

\section{Considerations in the Lattice Calculation}

{\it Choosing the lattice parameters.}
HQEFT becomes valid when the heavy quark mass
$m_Q \gg \Lambda_{QCD}$. Therefore we need to choose the lattice parameters
to satisfy this constraint.
For example,
at $\beta = 6.0$ the inverse lattice spacing $a^{-1} \approx 2.0$ GeV.
If we take
the physical value $\Lambda_{QCD} \approx 0.2$ GeV, then in lattice units
$\Lambda_{QCD} = 0.1$.
We need to choose the heavy quark mass $a m_Q$ to be much larger than $0.1$.
However, one would expect large lattice spacing artifacts for
$am_Q \raise.3ex\hbox{$>$\kern-.75em\lower1ex\hbox{$\sim$}} 1$.
At $\beta=6.0$ this means that the physical mass
of the heavy quark is limited to be in the range of the $D$ meson or smaller.
At larger $\beta$
values, the lattice spacing is
smaller, and we can  therefore
accommodate mesons
with larger physical masses on the lattice. In this work we choose
the heavy quark mass in the range $1.5-3$ GeV by tuning the
hopping parameter $\kappa_Q$ for the heavy quark Q.
On the other hand, the light quark $q$ in the heavy meson should have a mass
$m_q \leq \Lambda_{QCD}$. This can be achieved by changing the hopping
parameter
$\kappa_q$ for the light quark
and then extrapolating to the chiral limit $\kappa_{q,cr}$,
where the extrapolated pion mass is approximately matched to the physical pion
mass $m_\pi = 140$ MeV.

{\it How far can $v\cdot v^\prime$ change on the lattice?}
In the lattice calculations for the matrix element, we always have either
the initial or the final particle at rest. Thus
\begin{equation}
v\cdot v^\prime = \left\{ \begin{array}{ll}
                          {E_D / m_D} & \mbox{if $v^\prime= (0,0,0,1)$}\\
                          {E_D^\prime / m_D} & \mbox{if $v= (0,0,0,1)$}
                          \end{array} \right.~,
\end{equation}
with $E_D = \sqrt{ m_D^2 + {\vec p}^2} $.
For a spatial lattice $L = 24$, we have injected momenta
\begin{equation}
{\vec p} = {2\pi \over L} (1,0,0) ~, {2\pi \over L} (1,1,0) ~.
\end{equation}
Since $m_D \approx 1.0$ in lattice units, one gets $v \cdot v^\prime \approx
1.034$ and $1.066$ respectively. To get larger values for $v \cdot v^\prime$
one needs to inject larger lattice momenta which would in turn introduce large
statistical noise in the matrix element calculations. Thus
$v \cdot v^\prime$ can not be much larger than $\sim 1.1$
in current simulations.

{\it Removing the lattice artifacts; normalization.}
Since the heavy meson mass is near
1 in lattice units
the lattice artifacts could be significant. Ultimately the lattice artifacts
can be brought under control either by comparing data at different $\beta$
values or
by
using
improved lattice actions. However, for simulations at
a given $\beta$ value there are several ways to check
the size of the lattice artifacts.
Also, for the purpose of this work, we are able to get rid of most of the
undesirable effects by imposing the conservation of heavy quark flavor current,
as we will explain below.

For the elastic scattering process the matrix element has
the simple form given
in Eq.\ (\ref{eq:forwd}). This is derived from
Eq.\ (\ref{eq:form1}) which
is based on the assumption
that the theory is Lorentz invariant.
The lattice theory, however, is not exactly Euclidean
rotational invariant due to the finite lattice spacing $a$.
Thus $f_-$ is not exactly zero. The amplitude of $f_-$ (or $f_-/f_+$) gives
a measure for the violation of the Euclidean invariance on the lattice.

We can also estimate the size of the lattice artifacts by checking the
simulation
results against known continuum matrix element values at some special points.
For example, when both the initial and the final D mesons are at rest,
$v = v^\prime =
(0,0,0,1)$, the continuum matrix elements of ${\bar c}\gamma_4 c$ is known
because of the quark flavor current conservation
\begin{equation}
<D| {\bar c}\gamma_4 c| D> = 2 m_D ~.
\label{eq:norm}
\end{equation}
At this so-called ``zero-recoil point" we have
\begin{equation}
\xi_0 (1) = 1 ~.
\end{equation}
Both Eqs.\ (\ref{eq:matrix})
and (\ref{eq:norm}) will have $O(a)$ corrections
on the lattice which can come from different origins.

Part of the $O(a)$ effect can be disposed of, to
a good approximation, by
including a normalization factor
\begin{equation}
<\psi(x){\bar \psi}(0)>_{cont} = 2\kappa u_0 e^{m}<\psi(x){\bar
\psi}(0)>_{latt}~,
\end{equation}
where \cite{Lepage}
\begin{equation}
e^m = 1 + {1\over u_0}\left({1\over 2\kappa} - {1\over 2\kappa_{cr}}\right)~,
\end{equation}
with $u_0$ the ``tadpole improvement'' factor.

Another correction comes from the use of (nonconserved) local
vector current $V_\nu = {\bar c} \gamma_\nu c$ on the lattice \cite{karsten}.
This effect can be corrected by introducing a rescaling factor $Z_V^{loc}$
in the vector current. In perturbation theory $Z_V^{loc}$
is calculated to be \cite{Martinelli}
\begin{equation}
Z_V^{loc} = 1 - 27.5{g^2\over 16 \pi^2}.
\label{eq:Z_V}
\end{equation}

In general the size of lattice artifacts will be momentum dependent so the
$O(a)$ correction will be different for Eq.\ (\ref{eq:norm})
and Eq.\ (\ref{eq:forwd}).
However, since we are using only small momentum injections,
we assume that
the leading $O(a)$ correction is approximately independent of
$p, p^\prime$ and the Lorentz index $\nu$.
Thus by enforcing
Eq.\ (\ref{eq:norm}) as the normalization condition for the matrix element
$<D(v^\prime)|V_\nu|D(v)>$,
corrections due to both the $exp(m)$ factors and $Z_V^{loc}$ factors
are taken care of automatically (together with any other multiplicative
$O(a)$
factors).

{\it Corrections due to finite heavy quark mass}.
At what mass scale the heavy quark limit sets in is an important question.
The charm quark mass is not that much larger than $\Lambda_{QCD}$. If we
choose the
$\rho$-meson mass $m_\rho$ as the typical scale for QCD, it becomes
even less clear if $m_c$ is heavy enough.
In lattice calculations of the matrix elements
the size of the corrections to the symmetry limit ({\it i.e.}, $m_Q \rightarrow
\infty$) have been observed to be process dependent.
For example,
the heavy meson decay constant $f_P$  is found to have
a mass correction that is as large as $50\%$ at $D$-meson mass
scale \cite{Labrenz,Abada};
while the form factor $A_1(q^2)$ at the zero-recoil point
$q^2 = q^2_{max}$ for $D \to K^*$ decay shows very little mass dependence
even for fairly small values of the heavy
quark mass \cite{BEKS}.
Both observations are
consistent with theoretical expectations.
In HQEFT the leading correction due to the heavy quark mass is in general
$O(1/m_Q)$, as is the case for $f_P$. However, in the case
of $D \to K^*$,
the $O(1/m_Q)$ correction vanishes exactly
at the zero-recoil point and the finite mass correction there is at most
$O(1/m_Q^2) $\cite{Luke}. Thus,
in lattice calculations, the observations of
the linear $O(1/m_Q)$ correction for $f_P$ calculation
and little mass dependence of $A_1(q^2_{max})$ for $D \to K^*$ decay
seem to
indicate that the finite mass
corrections are under control within HQEFT at a scale $\sim m_c$ and
that
numerical simulation results in this range can be used and extrapolated
reliably to obtain the heavy quark limit results.

In principle, for data obtained at fixed masses one may estimate the size of
$O(1/m_Q)$ corrections in the following way:
the matrix element can be in general written as
\begin{equation}
<A(v^\prime)|V_\nu|B(v)> = \sqrt{m_A m_B} C_{AB} \left[ {\tilde f}_+
(v\cdot v^\prime)
(v^\prime + v)_\nu + {\tilde f}_-(v\cdot v^\prime)(v^\prime - v)_\nu\right]~.
\end{equation}
This parametrization is equivalent to the conventional parametrization
Eq.\ (\ref{eq:form1}); we have relations
\begin{equation}
{\tilde f}_+ = {1\over 2C_{AB}} \left[\sqrt{m_A \over m_B} (f_+ + f_-)
+ \sqrt{m_B\over m_A}(f_+ - f_-)\right] ~,
\end{equation}
\begin{equation}
{\tilde f}_- = {1\over 2C_{AB}} \left[\sqrt{m_A \over m_B} (f_+ + f_-)
- \sqrt{m_B\over m_A}(f_+ - f_-)\right] ~,
\end{equation}
In the $m_Q \to \infty$ limit
because of the constraints on the effective heavy quark fields
\begin{equation}
\gamma_\nu v_\nu h_v = h_v, \ \ \ {\bar h}_v \gamma_\nu v_\nu = {\bar h}_v~,
\end{equation}
we have \cite{Wise}
\begin{equation}
{\tilde f}_- = 0 ~.
\end{equation}
At finite $m_Q$, ${\tilde f}_- $ is nonvanishing.
Thus the value of ${\tilde f_-}$ or the ratio
\begin{equation}
{{\tilde f_-} \over {\tilde f_+}} = {1 + f_-/f_+ - {m_B\over m_A}(1 - f_-/f_+)
\over 1 + f_-/f_+ + {m_B\over m_A}(1 - f_-/f_+)} ~,
\label{eq:m_Q}
\end{equation}
gives a measure of the $O(1/m_Q)$ correction to the leading order HQEFT.
Unfortunately, for the elastic scattering process considered in this paper
this relation is
not useful because in this case we have $m_A=m_B$ so that
${\tilde f}_-/{\tilde f}_+ = f_-/f_+ = 0$ in the continuum.
However, in principle,
equation (\ref{eq:m_Q}) may be used to check the finite
mass correction in computations of $B \to D$ decay.

\section{Numerical results}

We use the standard Wilson action for
the propagation of both heavy and light quarks and work in the
quenched approximation.
The gauge configurations used are listed in Table 1.
The techniques for measuring the two-point
function and the three-point matrix elements are standard \cite{Bernard,ELC}.
Here we only list the results for the measured $m_D$, $f_+(0)$ and $f_+(q^2)$
in
Tables 2-7. To increase the statistics, we have
used symmetry properties of the Green functions and averaged over $\pm t$
and $\pm {\vec p}$ whenever possible.

{\it Estimating the size of the lattice artifacts}. Data for $f_+(0)$ and
$f_+(q^2)$ listed in Tables 2-7
are direct numerical results. They include
all the lattice artifacts. For example, comparing to the known continuum value
$f_+(0) = 1$, we observe that the lattice artifacts at $\beta=6.0$
are typically $20\%-40\%$
at $\kappa_Q = 0.118$ and less than $10\%$ at $\kappa_Q=0.135$.
\null From the computation of the ratio $f_-/f_+$
we find the violation of Euclidean invariance is typically $5\%-15\%$
for $\kappa_Q=0.118$ and $3\%-10\%$ for $\kappa_Q=0.135$.
Note that the reduction of
the lattice artifacts when $\kappa_Q$ is changed from $0.118$ to $0.135$ agrees
with our intuitive expectations that
those lattice artifacts that are due to the heavy quark mass approaching
the lattice ultra-violet cutoff should decrease
with a decrease in mass of the heavy quark.
One may try to use the factors $Z_V^{loc}$
and $e^{m}$ to remove
some of the artifacts.
At $\beta=6.0$
we get from
the perturbative calculation, Eq.\ (\ref{eq:Z_V}),
that $Z_V^{loc} \approx 0.7$
(using the ``boosted" effective gauge coupling ${\tilde g}^2 \approx 1.7$
as suggested in ref \cite{Lepage-perth}). The factor $e^{m}$ is about 2 for
$\kappa_Q = .118$ and 1.5 for $\kappa_Q=.135$. Including both these factors
the corrected $f_+(0)$ become around one within errors. This is in agreement
with other observations \cite{Labrenz} that $Z_V^{loc}$ and $e^{m}$ factors
seem to account for the largest part of the lattice artifacts.

{\it Normalization and the Isgur-Wise function}.
As explained in Section 2,
we define $\xi_0 (v\cdot v^\prime) = f_+(q^2)/f_+(0)$
in an effort to correct for possible lattice artifacts.
We list the
results in the last two columns
of Tables 2--7 \footnote{We have also computations at $\beta =5.7$ with
results
basically consistent with data listed here within large errors.
However, the heavy quark mass used (corresponding to
$\kappa=0.094$) is extremely high for that $\beta$
with potentially very large lattice artifacts. They are
therefore not used in this work.}.
This method for deducing $\xi_0$ has the significant advantage
that it is free of uncertainties in factors such as $Z_V^{loc}$
and $e^{m}$.
Note that in Eq.\ (\ref{eq:fplusxi}) there
is a factor $C_{cc}$ in the connection between $f_+$ and $\xi_0$.
This factor comes from integrating out the QCD effects from the heavy quark
scale down to a light scale $\mu$.
For the current calculation, however,
the heavy quark and hard gluon effects in
QCD are included dynamically,
and we are therefore determining the
combination $C_{cc} \xi_0$ at arbitrary scale $\mu$.  For simplicity
we choose $\mu=m_D$ and set $C_{cc}=1$ according to Eq.\ (\ref{eq:cfactor}).

{\it Finite volume effects and the dependence on the light quark mass}.
In an effort to keep finite volume effects under control,
we first study our data with light quark mass ($m_q$)
in the range of the s-quark mass ($m_s$).
On the lattice we choose
$\kappa = .154$ at $\beta=6.0$
and $\kappa=0.149$ at $\beta=6.3$. The physical meson
masses at these
$\beta$ and $\kappa$ values are in the range of $0.5$--$1$ GeV, corresponding
to bound states of ${\bar s} s$.
The results for the Isgur-Wise function are plotted in
Fig.~1.
All data seem to fall on a smooth curve.
In particular, data for the $16^3$ and $24^3$
lattices at $\beta=6.0$ agree
to a very good approximation ({\it i.e.}, well within statistical errors )
assuring us that in this limit finite size effects on these
lattices are small.
Since the physical volume of the $24^3$ lattice at $\beta = 6.3$
is the same as the physical volume of the $16^3$ lattice at
$\beta=6.0$,
we expect the finite size effects on the 6.3 lattice
also to be small.

As discussed in Section 2, for the elastic scattering of physical
$D$ mesons,
one should extrapolate the light quark hopping
parameter to the chiral limit $\kappa \to \kappa_{cr}$.
By taking this limit on a finite lattice, we expect not only
a
shift of the Isgur-Wise function due to the change of the
light quark mass, but also
a finite volume effect.
The former effect is
physical
but the latter one
is due to the limitations of a finite lattice size and should be brought
under control by carefully comparing results on different lattice sizes.
In practice, however, it is hard with a limited set of data
to separate the light quark mass dependence
from the finite size effect.
Since, as Table 8 shows,
the physical linear size of all our lattices are roughly in the range of
($100$ MeV$)^{-1}$, as the light quark mass gets smaller
than $m_s$ we expect finite size effects to become
important, especially on the two smaller lattices.
Indeed,
comparison of Fig.~2 with Fig.~1 shows that
all the data is shifted up in the chiral limit.
The size of shift is the smallest on the $\beta=6.0, 24^3$ lattice,
which has the largest physical volume.
Therefore, the $\beta=6.0,24^3$ lattice is the most suitable one
for extracting the dependence of the Isgur-Wise function on the
light quark mass.
Comparison of the numbers in
Tables 4 and 5 corresponding to $\kappa=0.154$ with those from the
extrapolated case ({\it i.e.}, for $\kappa=0.157$) indicates that
the form factors and the ratio $f_+/f_+(0)$ change
very little ({\it i.e.}, $\le 2\%$) at a fixed momentum $\vec p$.
Most of the change in the Isgur-Wise function comes from the change
of the D meson mass
(with the light quark mass)
which in turn shifts the value of $v \cdot v^\prime$.
We note that the light quark mass dependence of the Isgur-Wise
function has also been studied
using chiral perturbation theory
\cite{Jenkins} and model calculations \cite{Sadzi}.
It was found that
the shift due to the light quark mass is small:
at $y=1.2$ the shift in $\xi_0$ ranges from
$\approx -2 \%$ to about $7\%$.

{\it Comparison with model calculations and experimental data}.
As discussed in the previous two paragraphs,
we quote the $m_q \sim m_s$
data as our final results. They are listed in
Tables 2--7 (At $\beta=6.0, \kappa_q=0.154$ on the $16^3$ and $24^3$ lattices
and at $\beta=6.3, \kappa_q=0.149$ on the $24^3$ lattice) and plotted in
Fig.~1. For comparison
we also plotted the theoretical upper bound (dashed line) on the Isgur-Wise
function
derived from algebraic sum rules by Bjorken \cite{Bjorken}.
The lower curve (dotted line) is the lower
bound\footnote{Recent work
brings into question the validity of this lower bound
for the Isgur-Wise function \cite{Falk}. However,
one may take the result of Ref.\ \cite{Rafael} as a model calculation.}
derived in Ref. \cite{Rafael}.
All of our data appear to be, within errors, below the upper bound of Bjorken.
The solid curve going through the data points is a fit to the form
\cite{NeuRie}
\begin{equation}
\xi_0(y) = {2\over y+1} \exp\left[-(2\rho^2_{NR}-1){y-1\over y+1}\right]~,
\ \ \ y = v\cdot v^\prime ~.
\label{eq:neu-rie}
\end{equation}
Taking into account the correlations in the data, we find,
after fitting,
$\rho^2_{NR} = 1.41(19)$ with
$\chi^2/dof \approx 13/12$.
A similar fit to the
data (which, as explained in the preceding
paragraph, is in this case only from the
$\beta=6.0$,
$24^3$ lattice)
in the chiral limit
gives $\rho^2_{NR} = 1.09(28)$ with $\chi^2/dof \approx 1.4/4$.
However,  this shift between
the central values
(1.41 {\it vs.} 1.09) cannot be taken
as a reliable reflection of the light quark mass dependence of the Isgur-Wise
function. Indeed, holding $\rho^2_{NR}=1.41$ fixed
also provides a good fit ( with $\chi^2 \approx 2.6/4$ ) to the
$\beta=6.0, 24^3$ data in the chiral limit. Thus it is reasonable
to assume that the shift of the Isgur-Wise function from $m_q = m_s$ to the
chiral limit is
within the quoted statistical errors. So far as
other systematic effects such as the residual
$O(a)$ effects and $O(1/m_Q)$ effects are
concerned, we expect them to be rather small, since
we are able to get a good universal fit to all the data.
We therefore estimate the total systematic error to be at
most the same size as the statistical error,  and quote
$\rho^2_{NR} = 1.41\pm .19 \pm .19 $
as our
result
for $m_q \sim m_s$
as well as in the chiral
limit.

Close to $v \cdot v^\prime = 1$ the Isgur-Wise function can
be parametrized as
\begin{equation}
\xi_0 (v\cdot v^\prime) = 1 - \rho^2 (v\cdot v^\prime - 1)
+ O((v\cdot v^\prime-1)^2) ~.
\end{equation}
If we fit our data points to this linear form for $v \cdot v^\prime < 1.06$
we get $\rho^2 = 1.24(26)$ with $\chi^2/dof = 2.6/5$.
Once again, that value of $\rho^2$ also provides an
acceptable fit ($\chi^2 \approx 1.8/3$) for the three
data points with $v \cdot v^\prime < 1.06$
(of the $\beta=6.0$, $24^3$ lattice) in the chiral limit.
Therefore for the linear fit we quote the result as
$\rho^2 = 1.24\pm .26 \pm .26 $.
We list our linear fit result in Table 9 along with $\rho^2$ values
estimated by other authors.
We note that lattice and various model calculations seem
to be in rough agreement within errors.
However, the value of
Blok and Shifman tends to be
rather low compared to the lattice result.

In Fig.~3 we plot our lattice results taken from Fig.~1
against experimental data from ARGUS \cite{ARGUS}.
We have normalized the experimental graph with respect to $|V_{cb}|
\sqrt{\tau_B/(1.32\ {\rm psec.})} = 0.047$.
For comparison we also plotted the theoretical
upper bound \cite{Bjorken} (dashed line), our fit to Eq.\ (\ref{eq:neu-rie})
(solid line) and the curve obtained in Ref. \cite{Rafael} (dotted line).
It is gratifying
that the lattice data has much smaller errors than
the experimental results in the region $v \cdot v^\prime < 1.15$.

\section{Conclusions}

We have computed the Isgur-Wise function on the lattice from the
elastic scattering matrix element.
The calculation is done using the standard Wilson fermion formulation for both
heavy and light quarks in the quenched approximation.
We find that all of our numerical results are within the theoretical
upper bound\cite{Bjorken}.
The Isgur-Wise function ($\xi_0$) is deduced by taking
ratios of form factors.
In this approach the lattice artifacts
appear to be under control.
Although the heavy meson mass in our calculation is typically
order one in lattice units,
the residual $O(a)$ effects are
apparently small,
as indicated  by
direct comparison of the data at $\beta=6.0$ and $\beta=6.3$.
Further checks can also be done with an improved action scheme \cite{UKQCD}.
The lattice meson mass we used is in the range of physical D meson.
Thus there are potential $O(1/m_Q)$ corrections. Within the limited
range accessible in our simulations we have not observed strong
$m_Q$ dependence.

Our best lattice results are obtained
when
the light quark mass is set at the scale of s-quark mass.
Data from our largest lattice seems to indicate that the
Isgur-Wise function changes by less than our statistical
error between
$m_q \sim m_s$
and the
chiral limit.
A model independent
linear fit using data close to $v \cdot v^\prime =1$
gives the slope $\rho^2 = 1.24 \pm .26 \pm .26$.
Comparison of the slope $\rho^2$ with continuum model calculations
are given in Table 9.
A direct comparison to the ARGUS experimental results \cite{ARGUS}
is given in Fig.~3.
Our results are also in good agreement with the preliminary results of
UKQCD Collaboration \cite{UKQCD}.

\noindent{\it Acknowledgements}

We thank the
UKQCD Collaboration for sending us their preliminary results before
publication.
We would like to thank Jim Labrenz and Jim Simone for useful
conversations.

C.B. was partially supported by the DOE under grant number
DE2FG02-91ER40628.
Y.S. was supported in part under DOE contract DE-FG02-91ER40676 and NSF
contract PHY-9057173, and by funds from the Texas National Research Laboratory
Commission under grant RGFY92B6.
A.S. was supported in part by the DOE under grant
number DE-AC0276CH00016.

The computing for this project was done at the National
Energy Research Supercomputer Center in part under the
``Grand Challenge'' program and at the San Diego Supercomputer Center.

\pagebreak

\pagebreak
\begin{center}
\begin{tabular}{||l|l|l|l||} \hline
$\beta $ & $L^3T$ & \# conf.  & fitting range  \\ \hline
6.0  & $16^3 39$  & 19 & $10< t < 15$   \\ \hline
6.0  & $24^3 39$  & 8  & $10< t < 15$   \\ \hline
6.3  & $24^3 61$  & 20 & $15< t < 25$   \\ \hline
\end{tabular}
\end{center}
\vspace{0.2cm}
\noindent{\bf Table 1}: List of gauge configurations used in this work.

\vspace{1.0cm}
% kappa = 0.118, 16**3x39
\begin{center}
\begin{tabular}{||l|l|l|l|l|l|l||} \hline
$\kappa_q $ & $m_D$ & $f_+(0)$  & ${\vec p}$ & $f_+$ & $v \cdot v^\prime$
& $\xi_0$ \\ \hline
 0.152 &  1.237(6)  &0.69(12) & $(1,0,0)$ & 0.65(12) & 1.0408(4) & 0.953(22) \\
       &            &         & $(1,1,0)$ & 0.62(14) & 1.0790(8) & 0.911(57) \\
\hline
 0.154 &  1.202(8)  &0.74(15) & $(1,0,0)$ & 0.72(16) & 1.0434(5) & 0.974(30) \\
       &            &         & $(1,1,0)$ & 0.70(18) & 1.0842(10)& 0.945(79) \\
\hline
 0.155 &  1.186(8)  &0.77(19) & $(1,0,0)$ & 0.76(21) & 1.0445(6) & 0.983(42) \\
       &            &         & $(1,1,0)$ & 0.74(24) & 1.0869(11)& 0.95(11) \\
\hline
 $\kappa_{cr}$ &  1.151(8)  &0.82(19) & $(1,0,0)$ & 0.82(21) & 1.0567(8) &
1.00(4) \\
       &            &         & $(1,1,0)$ & 0.81(24) & 1.1103(15)& 0.99(11) \\
\hline
\end{tabular}
\end{center}
\vspace{0.2cm}
\noindent {\bf Table 2}: Data and extrapolation to
$\kappa_{cr}$ on the $\beta=6.0, 16^3\times39$ lattice.
The heavy quark hopping parameter is set to $\kappa_Q = 0.118$.
$\kappa_{cr} \approx 0.157$.

\vspace{1.0cm}
% kappa = 0.135, 16**3x39
\begin{center}
\begin{tabular}{||l|l|l|l|l|l|l||} \hline
$\kappa_q $ & $m_D$ & $f_+(0)$  & ${\vec p}$ & $f_+$ & $v \cdot v^\prime$
& $\xi_0$ \\ \hline
 0.152 &  0.885(6)  &1.00(12) & $(1,0,0)$ & 0.89(12) & 1.085(1) & 0.883(29) \\
       &            &         & $(1,1,0)$ & 0.77(14) & 1.160(2) & 0.771(72) \\
\hline
 0.154 &  0.847(7)  &1.03(15) & $(1,0,0)$ & 0.92(16) & 1.0937(14)& 0.894(40) \\
       &            &         & $(1,1,0)$ & 0.79(18) & 1.1749(25)& 0.77(9) \\
\hline
 0.155 &  0.829(7)  &1.06(19) & $(1,0,0)$ & 0.94(20) & 1.0980(15)& 0.894(40) \\
       &            &         & $(1,1,0)$ & 0.78(22) & 1.1841(27)& 0.74(12) \\
\hline
 $\kappa_{cr}$ &  0.791(7)  &1.08(19) & $(1,0,0)$ & 0.98(20) & 1.116(2) &
0.90(5) \\
       &            &         & $(1,1,0)$ & 0.79(22) & 1.222(4) & 0.73(12) \\
\hline
\end{tabular}
\end{center}
\vspace{0.2cm}
\noindent {\bf Table 3}: Data and extrapolation to
$\kappa_{cr}$ on the $\beta=6.0, 16^3\times39$ lattice.
The heavy quark hopping parameter is set to $\kappa_Q = 0.135$.
$\kappa_{cr} \approx 0.157$.

\pagebreak

\vspace{1.0cm}
% kappa = 0.118, 24**3x39
\begin{center}
\begin{tabular}{||l|l|l|l|l|l|l||} \hline
$\kappa_q $ & $m_D$ & $f_+(0)$  & ${\vec p}$ & $f_+$ & $v \cdot v^\prime$
& $\xi_0$ \\ \hline
 0.152 &  1.241(8)  &0.61(13) & $(1,0,0)$ & 0.60(14) & 1.0185(2) & 0.973(20) \\
       &            &         & $(1,1,0)$ & 0.57(14) & 1.0366(4) & 0.934(35) \\
\hline
 0.154 &  1.202(10) &0.64(13) & $(1,0,0)$ & 0.63(14) & 1.0201(3) & 0.975(20) \\
       &            &         & $(1,1,0)$ & 0.60(14) & 1.0396(7) & 0.938(42)
\\ \hline
 0.155 &  1.182(11) &0.66(12) & $(1,0,0)$ & 0.64(12) & 1.0209(4) & 0.973(20) \\
       &            &         & $(1,1,0)$ & 0.62(12) & 1.0412(8) & 0.936(49) \\
\hline
 $\kappa_{cr}$ &  1.144(10) &0.69(13) & $(1,0,0)$ & 0.67(14) & 1.0258(4) &
0.98(2)  \\
       &            &         & $(1,1,0)$ & 0.65(14) & 1.0511(9) & 0.94(5) \\
\hline
\end{tabular}
\end{center}
\vspace{0.2cm}
\noindent {\bf Table 4}: Data and extrapolation to
$\kappa_{cr}$ on the $\beta=6.0, 24^3\times39$ lattice.
The heavy quark hopping parameter is set to $\kappa_Q = 0.118$.
$\kappa_{cr} \approx 0.157$.

\vspace{1.0cm}
% kappa = 0.135, 24**3x39
\begin{center}
\begin{tabular}{||l|l|l|l|l|l|l||} \hline
$\kappa_q $ & $m_D$ & $f_+(0)$  & ${\vec p}$ & $f_+$ & $v \cdot v^\prime$
& $\xi_0$ \\ \hline
 0.152 &  0.889(7)  &0.96(10) & $(1,0,0)$ & 0.90(10) & 1.0413(6) & 0.944(17) \\
       &            &         & $(1,1,0)$ & 0.84(10) & 1.0802(12)& 0.873(33) \\
\hline
 0.154 &  0.847(8)  &0.99(10) & $(1,0,0)$ & 0.94(10) & 1.0458(8)& 0.949(19) \\
       &            &         & $(1,1,0)$ & 0.87(11) & 1.089(3)  & 0.883(40) \\
\hline
 0.155 &  0.826(9)  &1.01(10) & $(1,0,0)$ & 0.96(10) & 1.0476(10)& 0.948(18) \\
       &            &         & $(1,1,0)$ & 0.90(11) & 1.093(5) & 0.884(43) \\
\hline
 $\kappa_{cr}$ &  0.784(9)  &1.05(10) & $(1,0,0)$ & 1.00(10) & 1.0543(12)&
0.95(2) \\
       &            &         & $(1,1,0)$ & 0.90(11) & 1.106(2)  & 0.89(4) \\
\hline
\end{tabular}
\end{center}
\vspace{0.2cm}
\noindent {\bf Table 5}: Data and extrapolation to
$\kappa_{cr}$ on the $\beta=6.0, 24^3\times 39$ lattice.
The heavy quark hopping parameter is set to $\kappa_Q = 0.135$.
$\kappa_{cr} \approx 0.157$.

\pagebreak

\vspace{1.0cm}
% beta=6.3, kappa = 0.125, 24**3x61
\begin{center}
\begin{tabular}{||l|l|l|l|l|l|l||} \hline
$\kappa_q $ & $m_D$ & $f_+(0)$  & ${\vec p}$ & $f_+$ & $v \cdot v^\prime$
& $\xi_0$ \\ \hline
 0.148 &  0.928(4) &0.74(8) & $(1,0,0)$ & 0.69(8) & 1.0350(3) & 0.929(14) \\
       &            &       & $(1,1,0)$ & 0.66(9) & 1.0708(5) & 0.89(4) \\
\hline
 0.149 &  0.906(4) &0.78(10)& $(1,0,0)$ & 0.73(10) & 1.0373(3) & 0.937(16) \\
       &            &       & $(1,1,0)$ & 0.71(11) & 1.0752(5) & 0.91(5) \\
\hline
 0.150 &  0.884(4) &0.86(13)& $(1,0,0)$ & 0.83(13) & 1.0401(3) & 0.960(23) \\
       &           &       & $(1,1,0)$ & 0.84(17) & 1.0805(7) & 0.98(9) \\
\hline
 0.1507&  0.868(9) &0.99(18)& $(1,0,0)$ & 0.99(19) & 1.0423(6) & 0.993(32) \\
       &            &       & $(1,1,0)$ & 1.05(27) & 1.0836(17)& 1.06(12) \\
\hline
 $\kappa_{cr}$&  0.848(9) &1.02(18)& $(1,0,0)$ & 1.04(27) & 1.0466(6) & 1.02(3)
\\
       &           &        & $(1,1,0)$ & 1.12(27) & 1.0912(12)& 1.12(12) \\
\hline
\end{tabular}
\end{center}
\vspace{0.2cm}
\noindent {\bf Table 6}: Data and extrapolation to
$\kappa_{cr}$ on the $\beta=6.3, 24^3\times 61$ lattice.
The heavy quark hopping parameter is set to $\kappa_Q = 0.125$.
$\kappa_{cr} \approx 0.1516$.

\vspace{1.0cm}
% kappa = 0.140, 24**3x61
\begin{center}
\begin{tabular}{||l|l|l|l|l|l|l||} \hline
$\kappa_q $ & $m_D$ & $f_+(0)$  & ${\vec p}$ & $f_+$ & $v \cdot v^\prime$
& $\xi_0$ \\ \hline
 0.148&  0.587(2) &1.16(7)& $(1,0,0)$ & 0.99(6)  & 1.0894(6) & 0.85(3) \\
       &          &       & $(1,1,0)$ & 0.86(12) & 1.1743(10)& 0.74(11) \\
\hline
 0.149&  0.562(3) &1.18(8)& $(1,0,0)$ & 1.05(8)  & 1.0994(11)& 0.89(3) \\
       &          &       & $(1,1,0)$ & 0.98(16) & 1.193(2)  & 0.78(12) \\
\hline
 0.150&  0.537(3) &1.22(12)& $(1,0,0)$ & 1.09(11)& 1.1128(14)& 0.89(5) \\
       &          &        & $(1,1,0)$ & 1.06(21)& 1.216(3)  & 0.87(19) \\
\hline
 0.1507&  0.519(4) &1.27(18)& $(1,0,0)$ & 1.20(17)& 1.125(2)& 0.95(6) \\
       &           &        & $(1,1,0)$ & 1.33(34)& 1.235(3)  & 1.05(28) \\
\hline
 $\kappa_{cr}$&  0.496(4) &1.31(18)& $(1,0,0)$ & 1.27(17)& 1.131(2)& 0.98(6) \\
       &           &        & $(1,1,0)$ & 1.48(34)& 1.249(4)  & 1.08(28) \\
\hline
\end{tabular}
\end{center}
\vspace{0.2cm}
\noindent {\bf Table 7}: Data and extrapolation to
$\kappa_{cr}$ on the $\beta=6.3, 24^3\times 61$ lattice.
The heavy quark hopping parameter is set to $\kappa_Q = 0.140$.
$\kappa_{cr} \approx 0.1516$.

\pagebreak

\vspace{1.0cm}
\begin{center}
\begin{tabular}{||l|l|l|l|l|l||} \hline
$\beta $ & $a^{-1}$ & $L$  & $(aL)^{-1}_{phys}$ & $\kappa_Q$ & $m_D$ \\ \hline
 6.0 &  2.0 GeV  & 16  & 125 MeV & 0.118 & 2.3 GeV  \\ \hline
 6.0 &  2.0 GeV  & 24  & 83 MeV & 0.118 & 2.3 GeV  \\ \hline
 6.0 &  2.0 GeV  & 24  & 83 MeV & 0.135 & 1.6 GeV  \\ \hline
 6.3 &  3.2 GeV  & 24  & 125 MeV & 0.125 & 2.7 GeV  \\ \hline
 6.3 &  3.2 GeV  & 24  & 125 MeV & 0.140 & 1.3 GeV  \\ \hline
\end{tabular}
\end{center}
\vspace{0.2cm}
\noindent {\bf Table 8}: Physical values of the lattice spacing $a$, lattice
volume $aL$ and ``D'' meson mass $m_D$.

\vspace{1.0cm}
\begin{center}
\begin{tabular}{||l|l||} \hline
Lattice & $1.24 \pm .26 \pm .26$ \\ \hline
Rosner \cite{rosner} & 1.59(43) \\ \hline
Neubert \cite{neubert} & 1.42(60) \\ \hline
Burdman \cite{burdman} & 1.08(10) \\ \hline
Jin, Huang and Dai \cite{Jin}& 1.05(20) \\ \hline
Blok and Shifman \cite{Blok} & 0.65(15) \\ \hline
Bjorken \cite{Bjorken}& $0.25 < \rho^2$ \\ \hline
de Rafael and Taron \cite{Rafael} & $\rho^2 < 1.42 $ \\ \hline
\end{tabular}

\end{center}
\vspace{0.2cm}
\noindent {\bf Table 9}: Comparison of the lattice calculation,
using a linear fit to $\xi_0$, with various model calculations for
$\rho^2$ at $v\cdot v^\prime = 1$.

\pagebreak
\section*{Figure Captions}

\noindent{\bf Fig.~1}: The Isgur-Wise function is plotted against $v\cdot
v^\prime$. The lattice data are obtained
for $m_q \sim m_s$.
The parameters are:
$\beta=6.0, \kappa_Q=0.118, \kappa_q=0.154$ on $16^3\times 39$ lattice
(open triangle),
$\beta=6.0, \kappa_Q=0.135, \kappa_q=0.154$ on $16^3\times 39$ lattice
(solid triangle),
$\beta=6.0, \kappa_Q=0.118, \kappa_q=0.154$ on $24^3\times 39$ lattice
(open circle),
$\beta=6.0, \kappa_Q=0.135, \kappa_q=0.154$ on $24^3\times 39$ lattice
(solid circle),
$\beta=6.3, \kappa_Q=0.125, \kappa_q=0.149$ on $24^3\times 61$ lattice
(open square), and
$\beta=6.3, \kappa_Q=0.140, \kappa_q=0.149$ on $24^3\times 61$ lattice
(solid square).
Only statistical errors are shown.
The horizontal errors on the data points are smaller than symbol sizes.
The dashed curve is the upper bound on the Isgur-Wise
function \cite{Bjorken}. The solid curve is a fit
of the lattice data to Eq.\ (\ref{eq:neu-rie}).
The dotted curve is derived in Ref. \cite{Rafael}.

\noindent{\bf Fig.~2}:
The data for the Isgur-Wise function obtained
after extrapolations to the chiral limit.
The parameters are:
$\beta=6.0, \kappa_Q=0.118$, on $16^3\times 39$ lattice
(open triangle),
$\beta=6.0, \kappa_Q=0.135$, on $16^3\times 39$ lattice
(solid triangle),
$\beta=6.0, \kappa_Q=0.118$, on $24^3\times 39$ lattice
(open circle),
$\beta=6.0, \kappa_Q=0.135$, on $24^3\times 39$ lattice
(solid circle),
$\beta=6.3, \kappa_Q=0.125$, on $24^3\times 61$ lattice
(open square), and
$\beta=6.3, \kappa_Q=0.140$, on $24^3\times 61$ lattice
(solid square).
The data points can be compared to the corresponding
ones
in Fig.~1.
For reference we have shown the upper bound (dashed curve) on the Isgur-Wise
function \cite{Bjorken} and the curve (dotted) from Ref. \cite{Rafael}.

\noindent{\bf Fig.~3}: Comparison with experimental data.
The open circles are ARGUS experiment data from Ref. {\protect \cite{ARGUS}}
(We have normalized the graph with respect to $|V_{cb}|
\sqrt{\tau_B /1.32ps} = 0.047$). The crosses
are the lattice results
from Fig.~1. The solid, dashed and dotted curves have the same meaning
as in Fig.~1.
The horizontal errors on the lattice data points are smaller than symbol sizes.
(Note that the experimental point around
$v \cdot v^\prime = 1.1$
is shifted slightly for clarity.)

\begin{thebibliography}{99}

\bibitem{Wise} M. Wise, in {\it Particle Physics, the
Factory Era}, B.\ Campbell {\it et al.} eds.,
(Proceedings
of the Sixth Lake Louise Winter Institute, 1991), World Scientific,
Singapore, 1991, p.\ 222;
H. Georgi,
in {\it Perspectives in the Standard Model},
R.K.\ Ellis, C.T. Hill, and J.D. Lykken, eds.,
(Proceedings of the Theoretical Advanced Study Institute,
Boulder, 1991), World Scientific,
Singapore, 1992, p.\ 589.


\bibitem{ogilvie} J.\ Mandula and M.\ Ogilvie, Phys.\ Rev.\ D {\bf 45}
(1992) 2183.

\bibitem{Mack} P.~B.~Mackenzie, Nucl. Phys. B (Proc. Suppl.) 30 (1993) 35;
C.~T.~Sachrajda,
Nucl. Phys. B (Proc. Suppl.) 30 (1993) 20.

\bibitem{BSS} C.~Bernard, Y.~Shen and A.~Soni,
Nucl. Phys. B (Proc. Suppl.) 30, (1993) 473.

\bibitem{Bernard} C. Bernard, A. El-Khadra and A. Soni,
Phys. Rev. D43 (1991) 2140.

\bibitem{ELC} M.~Crisafulli, G.~Martinelli, V.~Hill and C.~Sachrajda,
Phys. Lett. B223 (1989) 90.

\bibitem{Lepage}
P. Lepage, Nucl. Phys. B (Proc. Suppl.) 26, (1992) 45;
P.~Mackenzie,
Nucl. Phys. B (Proc. Suppl.) 30, (1993) 35.;
A.S.~Kronfeld,
Nucl. Phys. B (Proc. Suppl.) 30, (1993) 445.

\bibitem{karsten} L. H. Karsten and J. H. Smit, Nucl. Phys. B183, (1981) 103.

\bibitem{Martinelli} G. Martinelli and Y. C. Zhang, Phys. Lett. B123,
(1983) 433.

\bibitem{Labrenz} C. Bernard, J. Labrenz and A. Soni,
Nucl.\ Phys.\  B (Proc.\ Suppl.) 30 (1993) 465;
``A Lattice Computation of the Decay Constants of $B$ and $D$ Mesons,''
UW/PT-93-06, June, 1993.

\bibitem{Abada} A. Abada, et al.  Nucl. Phys. B (Proc. Suppl.) 26, (1992) 344.

\bibitem{BEKS} C. Bernard, A. X. El-Khadra and A. Soni, Nucl. Phys. B
(Proc. Suppl.) 26, (1992) 204; Phys. Rev. D47 (1993) 998.

\bibitem{Luke} M.~E.~Luke, Phys. Lett. B252, (1990) 447;
M.~Neubert, Phys. Lett. B264, (1991) 455.

\bibitem{Lepage-perth}
G.~Peter Lepage and Paul B. Mackenzie, FERMILAB-PUB-19/355-T (9/92)

\bibitem{Jenkins} E. Jenkins and M.~J.~Savage, Phys. Lett. B281
(1992) 331.

\bibitem{Sadzi} M. Sadzikowski and K.~Zalewski, preprint TPJU-9-93,
hep-ph/9304319, and references therein.

\bibitem{Bjorken} J. D. Bjorken, SLAC report SLAC-PUB-5278 (1990).

\bibitem{Rafael} E. de Rafael and J. Taron, Phys. Lett. B282,
(1992) 215.

\bibitem{Falk} A.~F.~Falk, M.~Luke and M.~B.~Wise, SLAC report
Phys. Lett. B299 (1993) 123;
B.~Grinstein and P.~F.~Mende, Phys. Lett. B299 (1993) 127.

\bibitem{NeuRie} M.~Neubert and V.~Rieckert, Nucl. Phys. B382 (1992) 97.

\bibitem{rosner} J. L. Rosner, Phys. Rev. D42, (1990) 3732.

\bibitem{neubert} M. Neubert, Phys. Lett. B264, (1991) 455.

\bibitem{burdman} G. Burdman, Phys. Lett. B284, (1992) 133.

\bibitem{Jin} H. Y. Jin, C. S. Huang and Y. B. Dai, Z. Phys. C56 (1992) 707.

\bibitem{Blok} B. Blok and M. Shifman, Phys. Rev. D47 (1993) 2949.

\bibitem{ARGUS} ARGUS Collaboration, preprint DESY-92-146.

\bibitem{UKQCD} UKQCD Collaboration, to be published.


\end{thebibliography}
\end{document}